\def\eps#1#2{{\includegraphics[scale=#2]{#1}}}

\def\sl{\slshape}

\newcommand{\ci}{\cite}
\newcommand{\lab}{\label}
\newcommand{\eq}{\eqref}
\def\bl{\renewcommand{\baselinestretch}} 
\def\cl{\centerline}
\def\bib#1{\bibitem[#1]{#1}}

\def\lp{\left(} 
\def\rp{\right)}
\def\lb{\left[}
\def\rb{\right]}
\def\la{\langle\, }

\def\ra{\,\rangle}

\def\0#1{{(#1)}}
\def\1#1{{\hat #1}}
\def\2#1{{\tilde #1}}
\def\3#1{{\boldsymbol#1}}
\def\4#1{{\mathbb#1}}
\def\5#1{{\cal#1}}
\def\6#1{_{\scriptscriptstyle#1}}
\def\7#1{{\bar#1}}
\def\8{\infty}
\def\9#1{^{\scriptscriptstyle#1}}
\def\/#1{{\bf#1}}
\def\;#1{{\breve#1}}

\def\bbh#1{{\mathbb{\widehat#1}}}

\def\bb#1{{\boldsymbol{\bar#1}}}
\def\bh#1{{\boldsymbol{\hat{#1}}}}
\def\bt#1{{\boldsymbol{\tilde#1}}}

\def\a{\alpha} 
\def\b{\beta} 
\def\c{\chi}
\def\d{\delta} 
\def\e{\varepsilon}

\def\i{\iota}

\def\l{\lambda} 
\def\m{\mu} 

\def\o{\omega} 
\def\p{\pi} 

\def\q{\theta}

\def\t{\tau}

\def\D{\Delta}

\def\L{\Lambda} 
 
\def\P{\Pi}

\def\hb#1{{\qq\text{#1}\qq}}
\def\bull{$\bullet\ $}
\def\db{{d\kern-.8ex {^-}}}

\def\={\equiv} 
\def\app{\approx} 
\def\Box{\square}
\def\cc#1{{{\mathbb C\hskip.5pt}^{#1}}}

\def\curl{\nabla\times} 

\def\div{\nabla\cdot }

\def\ie{{\it i.e., }}
  
\def\imp{\ \Rightarrow\ }
\def\i1#1{\int_{-\infty}^\infty d#1\, }

\def\ir{\int_{-\infty}^\infty}

\def\pl{\partial}
\def\qq{\quad}

\def\re{{\,\rm Re}\  }   
\def\rr#1{{{\mathbb R}^{#1}}}

\def\sgn{{\rm sgn \,}}
 
\def\sr{\sqrt}

\def\sv#1{\vskip#1ex}
  \def\since{\ \because\ }

\def\bx#1{{ \boxed{\ #1\ }}}

\def\frame#1#2{
    \cl{\vbox{\hrule height .3pt
    \hbox{\vrule width .3pt\kern 5pt
    \vbox{\kern 5pt
    \vbox{\hsize #1cm\noindent#2}
    \kern 5pt}
    \kern 5pt\vrule width .3pt}
    \hrule height 0pt depth.3pt}}}

\def\ss{\cdot\3s}
\def\oo{\cdot\3\o}

\def\xt{\cdot d\3x dt}

\def\*{*\!}

\def\ast{\1s\int_{C_s} d\2k\ e^{ikz}}
\def\as{\int_{C_s} d\2k\ e^{ikz}}

\documentclass[12pt]{article}  
\usepackage{amssymb,amsmath}  
\usepackage{graphicx}
  
\begin{document}

\title{\bf Electromagnetic Wavelets  as  Hertzian Pulsed Beams in Complex Spacetime}

\author{To Professor Jerzy Pleba\'nski on his 75th birthday\\ \\ 
Gerald Kaiser\\
   The Virginia Center for Signals and Waves\\
   kaiser@wavelets.com $\bullet$\  www.wavelets.com\\ 
}

\maketitle

\begin{abstract}\noindent
Electromagnetic wavelets  are a family of  $3\times3$ matrix fields
$\4W_z(x')$ parameterized by complex spacetime points
$z=x+iy$ with $y$ timelike. They are  translates of a \sl basic \rm wavelet
$\4W\0z$ holomorphic in the future-oriented union $\5T$ of the forward and
backward tubes. Applied to a  polarization vector $\3p=\3p_m-i\3p_e$,
$\4W\0z$ gives an anti-selfdual solution $\4W\0z\3p$  derived from a selfdual Hertz
potential $\bt Z\0z=-iS\0z\3p$, where $S$ is the \sl Synge function \rm  
acting as a  Whittaker-like scalar Hertz potential. Resolutions of unity
exist giving representations of sourceless electromagnetic fields as superpositions of
wavelets. With the choice of a branch cut,  $S$ splits into a difference
$S\9+\0z-S\9-\0z$ of retarded and advanced \sl pulsed beams \rm whose limits as
$y\to 0$ give the propagators of the wave equation.  This yields a similar splitting of
the wavelets and leads to their complete physical interpretation as pulsed beams
absorbed and emitted by a \sl disk source \rm $D\0y$ representing the branch cut.
The choice of $y$ determines the beam's orientation, collimation and duration, giving
beams as sharp with pulses as short as desired.  The sources are computed
as spacetime distributions of electric and magnetic dipoles supported on
$D\0y$.  The wavelet representation of sourceless electromagnetic fields now splits
into representations with advanced and retarded sources. These representations are the
electromagnetic counterpart of relativistic coherent-state representations  previously
derived for massive Klein-Gordon and Dirac particles.

\end{abstract}

\section{Hertz Potentials} 

Acoustic and electromagnetic wavelets were introduced in    
\ci{K92, K94, K94a} as \sl generalized frames \rm of localized solutions of the
homogeneous  wave equation and Maxwell's equations, giving wavelet-like
representations of general sourceless solutions. They were defined in  Fourier
space, and explicit space-time expressions were found for the acoustic  but not the 
electromagnetic wavelets.  Here we compute  the electromagnetic wavelets in
spacetime and show that they have a simple interpretation as pulsed beams emitted
and absorbed by localized electric and magnetic dipole distributions.
By separating their advanced and retarded parts, the wavelet representation of
electromagnetic fields is generalized to include localized sources, interpreted as
polarization distributions generating Whittaker-like Hertz potentials.

We begin by reviewing the derivation of electromagnetic  fields by Hertz
potentials. Our presentation is essentially a translation of \ci{N55} to the language of
differential forms \ci{AMT88, T96}.  An electromagnetic field is represented by two
2-forms
$F, G$ satisfying Maxwell's equations,
\begin{align}
dF&=0 \qq (F=\3E\xt+\3B\ss)\lab{F}\\
dG&=J \qq (G=\3D\ss-\3H\xt).\lab{G}
\end{align}
Here $J$ is the current 3-form, products of forms are wedge products,
$d\3x=(dx^1, dx^2, dx^3)$, and $\3s$ is the vector-valued spatial area form $\3s=-*
d\3x dt=(dx^2 dx^3, dx^3 dx^1, dx^1 dx^2)$, where $*$ is the Hodge duality
operator  in Minkowski space, defined
\ci{R} so that $**=-1$.

In addition to \eq{F} and \eq{G},  we need constitutive relations,  given here in
terms of the electric and magnetic polarization densities $\3P_e,
\3P_m$ of the medium:
\begin{align}\lab{P}
*F+G=P \qq (P=\3P_e\ss+\3P_m\xt),
\end{align}
where $*F=-\3E\ss+\3B\xt$ is the Hodge dual of $F$.  Without specifying initial and
boundary conditions, the best we can hope for is \sl general \rm (or \sl
local) \rm solutions up to an arbitrary sourceless field. We now show to derive these
using Hertz potentials.

Since $F$ is closed, it can be derived from a potential 1-form $A$ subject to gauge
transformations:
\begin{align}\lab{A}
F&=dA=dA', \qq A=A_\m dx^\m, \qq A'=A-d\c.
\end{align}
Applying the co-differential operator $\d=*d*$ gives
\begin{align}\lab{G1}
\d A'=\d A-\d d\c=\d A-\Box\c,
\end{align}
where $\Box=d\d+\d d$ is the (Hodge) d'Alembertian operator on differential forms, 
reducing to the  wave operator on functions. Locally at least, $\c$ can be chosen
to satisfy $\Box\c=\d A$, so that $\d A'=0$ and $A'$ is in the 
Lorenz gauge.\footnote{Due to L.V. Lorenz, not H.A.
Lorentz; see Penrose and Rindler \ci{PR84}.}  
By the Poincar\'e lemma, $*A'$ can be derived from a potential 2-form:
\begin{align*}
d*A'=0 \imp *A'=-dZ \qq (Z=\3Z_e\ss+\3Z_m\xt).
\end{align*}
This shows that the original potential, which need not be in Lorenz gauge, can be
written locally as
\begin{align}\lab{Z}
A=*dZ+d\c, \qq   F=d*dZ.
\end{align}
$Z$ is called a \sl Hertz potential \rm with electric and magnetic vectors $\3Z_e,
\3Z_m\,$.   The relation between the components of $F$ and $Z$ can be expressed
compactly in terms of the \sl anti-selfdual \rm form $F\9-=i*F\9-$ and the \sl selfdual
\rm form $Z\9+=-i*Z\9+$, defined by
\begin{align}
2F\9-=F+i*F&=2\3F\oo \qq 2\3F=\3E+i\3B \qq \3\o=d\3x dt-i\3s\notag\\
2Z\9+=Z-i*Z&=2\3Z\cdot\bb\o
\qq 2\3Z=\3Z_m-i\3Z_e \qq \bb\o=d\3x dt+i\3s.\lab{FZ}
\end{align}
Namely,
\begin{align*}
2F\9-&=d*dZ+i\d dZ=d*dZ-id\d Z+i\Box Z\\
&=d*dZ-id*d*Z+i\Box Z=2d*dZ\9++i\Box Z,
\end{align*}
which translates to the complex vector equation
\begin{align}\lab{FZ2}
2\3F=2i\5L\3Z-\Box\3Z_e
\end{align}
where $\5L$ is the operator
\begin{align*}
\5L\,\3Z\= \curl(\curl\3Z)+i\pl_t\curl\3Z.
\end{align*}
To obtain a wave equation for $Z$ in terms of the sources, we need  a \sl stream
potential \rm  \ci{N55} for $J$, \ie  a  2-form
$G^\star$ such that
\begin{align*}
J=-dG^\star \qq   (G^\star\=-\3D^\star\ss+\3H^\star\xt).
\end{align*}
The resemblance  $G^\star\sim -G$ is intentional, since one possibility is 
$G^\star=-G$. But this would be circular since the field is unknown.  A simple stream
potential $G^\star_0$ can be defined  as a time integral of the current.   Then $G$ 
differs from $-G^\star_0$ at most by a  homogeneous   solution of \eq{G}, \ie an
arbitrary exact 2-form:
\begin{align}\lab{da}
G=-G^\star_0+d\a,\qq \a\in\L^1.
\end{align}
The constitutive relation \eq{P} implies
\begin{align*}
P=*F+G=\d dZ-G^\star_0+d\a=\Box Z -d\d Z-G^\star_0+d\a,
\end{align*}
or
\begin{align}\lab{wave}
\bx{\Box Z=P+G^\star}
\end{align}
where
\begin{align*}
 G^\star=G^\star_0-d\b,\qq  \b=\a-\d Z
\end{align*}
is another stream potential. The dependence of $\b$ on $Z$ is not a
problem since $\b$, like $\a$, is  arbitrary.  The selfdual representation of
\eq{wave} is
\begin{align}\lab{ZP}
\Box \3Z=\3P+\3G^\star,\qq  2\3P=\3P_m-i\3P_e,
\qq  2\3G^\star=\3H^\star+i\3D^\star\,.
\end{align}

Thus $Z$ is produced by a unified source composed of  material polarization
and stream potential; the latter may be viewed as an \sl  effective  \rm classical 
vacuum polarization induced by the current.  If the medium is polarizable,
magnetizable or conducting, these sources in turn depend on the field. In that case it is
better  to use the common multiplicative form of the constitutive relations for the
induced sources (assuming linearity), which gives a modified version of \eq{wave}. For
a thorough study of Hertz potentials and their gauge theory, see \ci{N55, N57}.

\section{Hertz potentials in Fourier space}

Initially, the construction of electromagnetic wavelets will be based on holomorphic
fields obtained by extending \sl sourceless \rm anti-selfdual solutions to complex
spacetime. Eventually, sources will be introduced that preserve the holomorphy of
the fields \sl locally, \rm outside the sources. But for now, we specialize to
a homogeneous field in vacuum,
\begin{align}\lab{Max}
J=P=0\imp dF\9-=0\imp i\pl_t\3F=\curl\3F,\qq \div\3F=0.
\end{align}
Solutions are given in terms of selfdual Hertz potentials by \eq{FZ2}:
\begin{align}\lab{wave2}
\Box\3Z\0x=\30, \qq  \3F\0x=i\5L\3Z\0x.
\end{align}
The Fourier  solution of the wave equation is
\begin{align}\lab{FT}
\3Z\0x=\int_C d\2k \ e^{ikx} \bh Z\0k
=\int_{C\6+} d\2k \ e^{ikx} \bh Z\0k+\int_{C\6-} d\2k \ e^{ikx} \bh Z\0k,
\end{align}
where $x=(\3x, t),\  \ k=(\3k, k_0), \  \ kx=k_0t-\3k\cdot\3x,\ \  C\6\pm$ are the
positive and negative-frequency light cones
\begin{align*}
C\6\pm=\{k: \pm k_0=\o>0\},\  \  \o\=|\3k|, \hb{and} 
d\2k=d^3\3k/(16\p^3 \o)
\end{align*}
is the Lorentz-invariant measure on the double cone $ C=C\6+\cup C\6-$. 
The coefficient function $\bh F\0k$ is the restriction of the Fourier transform of
$\3F\0x$ to $C$.  Inserting the definition of $\5L$,
\begin{align}\lab{F2}
\3F\0x=\int_C d\2k \ e^{ikx}\bh F\0k=i\!\!\int_C d\2k \ e^{ikx}
\lb -\3k\times(\3k\times\bh Z) +ik_0\3k\times\bh Z\rb.
\end{align}
Note that \eq{Max} requires 
\begin{align}\lab{Max2}
\bh F=i\3n\times\bh F, \hb{where} \3n\0k=\3k/k_0, \qq \3n^2=1\hb{on} C.
\end{align}
Thus for every $k\in C$,  $\bh F\0k$ is an eigenvector with
eigenvalue 1 of
\begin{align*}
\4S\0k: \cc3\to\cc3 \hb{defined by} \4S\,\3v=i\3n\times\3v.
\end{align*}
But
\begin{align*}
\4S^2\3v=\3v-\3n(\3n\cdot\3v)\imp \4S^3=\4S,
\end{align*}
so $\4S$ has the nondegenerate spectrum $\{-1, 0, 1\}$ and the
orthogonal projection to the eigen\-space with $\4S=1$ is
\begin{align}\lab{PP}
\4P\0k=\frac{1}{2}\lb \4S^2+\4S \rb \imp \4S\4P=\4P=\4P^*=\4P^2.
\end{align}
We can now write \eq{F2} in the form
\begin{align}\lab{F3}
\3F\0x=\int_C d\2k \  e^{ikx}\bh F\0k,
\qq  \bh F=2i\o^2\4P\,\bh Z=\4P\,\bh F.
\end{align}

\section{Extension to complex spacetime}

Sourceless waves are ``boundary values'' of holomorphic functions in a sense to be
explained.   Denote the future and past cones at $x=0$ by
\begin{align*}
V\6\pm=\{y=(\3y, s)\in\rr4: \pm s>|\3y|\}.
\end{align*}
The \sl forward and backward tubes \rm \ci{SW64, PR84} are the complex domains
\begin{align*}
\5T\6\pm=\{x+iy\in\cc4: y\in V\6\pm\}.
\end{align*}
Since they are disjoint, we are free to give them independent orientations. We
orient them  \sl oppositely, \rm  considering $\5T\6+$ to have a \sl positive \rm and 
 $\5T\6-$ a \sl negative \rm orientation. Specifically, let $\5T\6+$ be oriented by
its volume form $\t=d^4x\, d^4 y$ and $\5T\6-$ by the \sl time-reversed \rm form
$-\t$ ($d\3x\to d\3x, dt\to -dt, dy^\m\to-dy^\m$). The oriented union (chain)
\begin{align*}
\5T=\5T\6+ -\5T\6-
\end{align*}
will be called the \sl causal tube. \rm  

We will extend sourceless fields such as $\3F\0x$ to $\5T$,  then interpret
these extensions $\bt F\0z$ physically by examining their behavior on real spacetimes
defined by
\sl slices \rm 
\begin{align*}
\4R^4_y=\{x+iy: x\in \rr4\},\qq y\in V\6\pm\,,
\end{align*}
in the spirit of Newman, Pleba\'nski, Penrose, Rindler, Robinson, Trautman  and others 
\ci{N65, NJ65, N73, NW74, N02, PR78, PR84, T62}. As $y\to 0$, the extended fields
converge to their ``boundary values''  on $\rr4$ in a sense to be made specific. (It is
only in \sl this \rm sense that  $\bt F\0z$ is to be regarded an extension of $\3F\0x$.)
But note that $\rr4$ is not the topological boundary of  $\5T\6\pm$, which would be
the 7-dimenional set $\{x+iy:
\pm y_0>0,\ y^2=0\}$, but its  \sl Shilov boundary \rm \ci{H73}. 

Solutions of homogeneous wave equations, such as $\3Z\0x$  in \eq{FT}, can be
extended to $\5T$ by the \sl analytic-signal transform\rm
\begin{align}\lab{ast}
\bt Z\0z=\1s\int_{C_s} d\2k \  e^{ikz} \bh Z\0k,\qq  z=x+iy\in\5T,
\end{align}
where
\begin{align*}
\1s=\sgn s,\qq  C_s=\begin{cases}C\6+ &\text{if }s>0 \\ 
 C\6-&\text{if } s<0. \end{cases}\qq  (s\=y_0).
\end{align*}
Note that \sl the positive and negative frequency parts of $\3Z$ are extended to the
forward and backward tubes, respectively. \rm Since $\3Z$ is complex to begin with,
these parts are independent.  The exponential factor $e^{-ky}$ in \eq{ast} decays as
$|\3k|\to\8$ whether $y\in V\6\pm\,$, so if $\bh Z\0k$ is reasonable (of polynomial
growth, say), then the integral defines $\bt Z\0z$ as a holomorphic function.  
Since $e^{-ky}$ decays \sl least \rm rapidly along those rays $k\in C\6\pm$ that are
``nearly parallel'' to $y\in V\6\pm$, it follows that such rays are \sl favored \rm
by the extension \eq{ast} if $y$ is ``nearly lightlike.''

Since $\3Z\0x$ is the \sl sum \rm of its positive and negative frequency parts, the
sign $\1s$ in \eq{ast} tells us that it can be recovered (in a distributional sense) as the
\sl difference \rm of the (Shilov) boundary values
\begin{gather}\lab{diff}
\3Z\0x=\bt  Z(x+i\e 0)- \bt Z(x-i\e0), 
\end{gather}
where 
\begin{align*}
\bt Z(x\pm i0)=\lim_{\e\searrow 0}\bt Z(x\pm i\e y),\qq y\in V\6+
\end{align*}
and the limit can be shown to be independent of $y\in V\6+$.  

It is also possible to define $\bt Z$ more suggestively as
\begin{align}\lab{ast2}
\bt Z(x+iy)=\frac{\1s}{2\p i}\ir \frac{d\t}{\t-i}\ \3Z(x+\t y),
\end{align}
which is defined (but not holomorphic) for all $y\in\rr4$ and can be shown to 
reduce to \eq{ast} when $y\in V\6\pm$\,. This definition was used in
\ci{K90, KS92, K94} \sl  without \rm the orienation  $\1s$. Only recently have I
understood the value of this orientation, which connects the transform to
hyperfunction theory and locality (see below).

The electromagnetic field can now be  extended similarly as
\begin{align}
\bt F\0z&=\1s\int_{C_s} d\2k \  e^{ikz} \bh F\0k=i\5L\bt Z\0z
=2i\1s\int_{C_\1s} d\2k \  e^{ikz} \o^2\4P\0k\bh Z\0k, \lab{F4}
\end{align}
where $\5L$ differentiates with respect to $x$ (so as to be defined in $\4R^4_y$).
The positive and negative-frequency parts of $\bh F$ also have positive
and negative \sl helicities \rm \ci{K94}, so  the restrictions of $\bt F\0z$ to $\5T\6+$
and
$\5T\6-$ are positive and negative-helicity solutions. 

In one dimension, the above extension is trivial: the positive and negative
frequency   parts extend to the upper and lower complex half-planes, and a quite
general distribution can be written in the form \eq{diff}. (An elementary form of this
idea was applied to communication theory by Dennis Gabor, who called
the extensions \sl analytic signals; \rm hence the name.) In dimension $n>1$, we need
something to replace the half-lines of  positive and  negative  frequencies. When the
support in Fourier space is contained in a double cone (like the convex hulls of
$C\6\pm$), these become the appropriate replacements and the tubes over the open
\sl dual \rm cones (in this case $V\6\pm$) replace the upper and lower half-planes.

A \sl hyperfunction \rm  $H\0x$ in $\rr n$ \ci{K88, KS99} is a 
generalized  distribution defined, roughly, by differences in the
boundary values of a set of holomorphic functions $\2H_k\0z$ in a set of complex
domains $\5D_k$ enveloping the support of $H$. The functions $\2H_k\0z$ are called
\sl generating functions \rm for $H\0x$.\footnote{More accurately, differences in
boundary values must in general be replaced by sheaf cohomology. Fortunately, for
solutions of homogeneous relativistic equations (Klein-Gordon, Dirac, etc.), simple
differences like \eq{diff} suffice. 
} 
Equation  \eq{diff} \sl suggests \rm thinking of the restrictions of
$\bt Z\0z$ and $\bt F\0z$ to $\5T\6\pm$ as generating functions for $\3Z\0x$ and
$\3F\0x$.  We will not attempt to make rigorous use of hyperfunction theory; rather,
it will serve mainly as a guide. This will not only free us from its highly technical
requirements but also allow us to go beyond it, as some of our constructions (like the
extended delta functions $\2\d^3(\3z)$ and $\2\d^4\0z$ below) do not fit neatly into
the theory. Also, since our goal is to understand physics directly  in
$\5T$,  we do not view $\bt F\0z$ and $\bt Z\0z$ merely as tools for the
analysis of the boundary distributions $\3F\0x$ and $\3Z\0x$,  as they would be in
hyperfunction theory.

\section{Electromagnetic wavelets}

To construct the wavelets, we need a Hilbert space of solutions.
Initially the inner product will be defined in  Fourier space.  It is uniquely determined
 (up to a constant factor) by the requirement of Lorentz invariance to be
\begin{align*}
\la\3F_1, \3F_2\ra=\int_C\frac{d\2k}{\o^2}\  \ \bh F_1^*\bh F_2
=4\int_Cd\2k\ \o^2\,\bh Z_1^*\4P\,\bh Z_2\,.
\end{align*}
Note that the measure $d\2k/\o^2\propto d^3\3k/\o^3$ is invariant under  scaling.
In fact, an equivalent inner product has been shown to be invariant under the   \sl
conformal group \rm $\5C$  of Minkowski space \ci{Gr64}. Therefore the Hilbert
space of anti-selfdual solutions
\begin{align*}
\5H=\{\3F: \|\3F\|^2=\la\3F,\3F\ra<\8\}
\end{align*}
carries a unitary representation of $\5C$.  We want to construct the wavelets so as
to preserve the connection with the conformal group. We therefore define the
wavelets in the same spirit as the \sl relativistic coherent states \rm
for the free Klein-Gordon and Dirac fields  \ci{K77, K78, K87, K90}.  For any \sl fixed
\rm  $z\in\5T$, consider the  \sl evaluation map \rm 
\begin{align*}
\5E_z: \5H\to\cc3\hb{defined by} \5E_z\3F=\bt F\0z.
\end{align*}
This is a bounded operator between Hilbert spaces (where $\cc3$ is given its standard
inner product). Its adjoint is, by definition, the 
electromagnetic wavelet\footnote{We are really using a vectorial version of the  Riesz
representation theorem. If $\5H$ were a Hilbert space of sufficiently nice \sl scalar \rm
functions, $\5E_z$ would be a bounded linear functional and $\5E^*_z(1)$ its
representation by the unique elelment of $\5H$ guaranteed to exist by the Riesz 
theorem. }
\begin{align}\lab{Wz}
\4W_z=\5E_z^*: \cc3\to\5H.
\end{align}
$\4W_z$ maps any \sl polarization vector \rm  $\3p\in\cc3$ to a
solution $\4W_z\3p\in\5H$. Therefore $\4W_z(x')$ must be a \sl matrix-valued \rm
solution of  Maxwell's equations (three columns, each a solution).   But
\begin{align*}
\4W_z^*\3F=\5E_z\3F=\bt F\0z=\ast\4P\bh F\0k\qq  \since\4P\bh F=\bh F,
\end{align*}
therefore, remembering the measure $d\2k/\o^2$ for the inner product in $\5H$,
the expressions for $\4W_z$ in the Fourier and spacetime domains are
\begin{align}\lab{Wz2}
\bbh W_z\0k =\1s\,e^{-ik\7z}\o^2\4P\0k,\qq 
\4W_z(x')=\1s\int_{C_s} d\2k\ e^{ik(x'-\7z)}\o^2\4P\0k.
\end{align}
The \sl reproducing kernel \rm  is defined as
\begin{align}\lab{RK}
\4K(z',\7z)=\4W_{z'}^*\4W_z=\q(y'y)\int_{C_s} d\2k\
e^{ik(z'-\7z)}\o^2\4P\0k\=\q(y'y)\4W(z'-\7z),
\end{align}
where $\q$ is the Heaviside step function, the factor $\q(y'y)$  enforces
the mutual orthogonality of wavelets parameterized by in the forward and backward
tubes, and the holomorphic matrix function
\begin{align}\lab{W}
\4W\0z=\as\o^2\4P\0k
\end{align}
generates the entire wavelet family by  translations: 
\begin{align}\lab{Wz3}
\4W_z(x')=\1s\4W(x'-\7z),\qq  z\in\5T.
\end{align}
We now compute $\4W\0z$ explicitly.  
Applying it to a vector $\3p\in\cc3$ gives
\begin{align}\lab{Wp}
2\4W\0z\3p=2\as\o^2\4P\0k\3p=i\5L\2R\0z\3p,
\end{align}
where
\begin{align}\lab{2R}
i\2R\0z=\as=\1s\int_{C_s} d\2k\ \1k_0 \,e^{ikz},\qq \1k_0=\sgn k_0\,.
\end{align}
The second equality, which holds because $s$ and $k_0$ have the same sign, shows
that $\2R\0z$ is the analytic-signal transform of
\begin{align}\lab{R}
R\0x=-i\int_Cd\2k\ \1k_0 \,e^{ikx}.
\end{align}
It is easily checked that this integral gives the (unique) solution to the follwing
inital-value problem:
\begin{align*}
\Box R\0x=0,\qq R(\3x, 0)=0,\qq \pl_tR(\3x, 0)=\d^3(\3x).
\end{align*}
$R$ is known as the \sl Riemann function \rm of the wave equation \ci{T96}.
The integral is readily found to be the difference between the \sl retarded
and  advanced  propagators, \rm 
\begin{align}\lab{Riem}
R\0x&=R\9+\0x-R\9-\0x,\qq R\9\pm\0x=\frac{\d(t\mp r)}{4\p r}.
\end{align}
On the other hand, we find
\begin{align}\lab{Synge}
S\0z\=i\2R\0z=\as=-\frac{1}{4\p^2z^2},\qq  z^2=z_\m z^\m=z_0^2-\3z^2.
\end{align}
This is nothing but  Synge's  ``elementary solution''   \ci[p.~360]{S65} of the wave
equation!  Among other things, it has been used by Trautman  \ci{T62}
to construct  interesting (null, curling) analytic solutions of the Maxwell and  linearized
Einstein equations.  

I believe this connection between the Riemann and Synge functions
confirms that the transform \eq{ast} is the ``right'' extension of
solutions to $\5T$.  In the absence of the orientation $\1s$ (with \eq{diff} now a
sum),  the Synge function is no longer the analytic signal of $iR\0x$. In
particular, we lose the connection with Huygens' principle in the limit $y\to 0$.

The Synge function is holomorphic on the complement of the complex \sl null cone
\rm
\begin{align*}
\5N=\{z\in\cc4: z^2=0\}.
\end{align*}
In particular, note that
\begin{align*}
x+iy\in\5N\cap\5T\imp x^2=y^2>0,\hb{and} xy=0.
\end{align*}
But the second equality is impossible since, by the first, $x$ and $y$ are both timelike.
Thus $S\0z$ is holomorphic in $\5T$.

According to \eq{wave2},  $2\4W\0z\3p$ can thus be
computed from the selfdual Hertz  potential 
\begin{align}\lab{Rp}
2\bt Z\0z=\2R\0z\3p,
\end{align}
which gives explicit expressions to the entire wavelet family.

\bull All the wavelets  are  translations of $\4W\0z$.
This can be further reduced by using the symmetries of $\4W\0z$, which reflect those
of Maxwell's equations --- \ie the conformal group.
For example,  $\4W\0z$ is homogeneous  of degree $-4$, so the  Lorentz norm of
$y$ can be interpreted as a \sl scale parameter: \rm
\begin{align}\lab{scale}
\4W_z(x')=\l^{-4}\,\4W\lp  (x'-\7z)/\l\rp, \qq  \l=\sr{y^2}\,>0.
\end{align}

\bull It is easily shown that
\begin{align*}
\4W\0z^*=\4W(\7z), \hb{hence} \4W_z(x')^*=\4W_{\7z}(x').
\end{align*}

\bull There exist (many equivalent) \sl resolutions of
unity \rm \ci{K94}, obtained by integrating over various subsets $\5D\subset\5T$
with appropriate measures $d\m_\5D$:
\begin{align}\lab{RU}
\int_\5D d\m_\5D\0z \4W_z \4W_z^*=I_\5H\,.
\end{align}
This is a ``completeness relation'' dual to the (non-) ``orthogonality'' relation
\eq{RK}. Each resolution gives a  representation  of solutions as
superpositions of wavelets,
\begin{align}\lab{RU2}
\3F(x')=\int_\5D  d\m_\5D\0z \4W_z(x')\4W_z^*\3F
=\int_\5D  d\m_\5D\0z \4W_z(x')\bt F\0z,
\end{align}
with the analytic signal $\bt F\0z$ on $\5D $ as the ``wavelet transform.''    One
natural subset for a resolution is the \sl Euclidean region \rm of real space and
imaginary time,
\begin{align*}
\5E=\{(\3x, is): \3x\in\rr3,\ s\ne 0\},\qq  d\m_\5E(\3x,  is)=d^3\3x\, ds.
\end{align*}
Since $\l=|s|$ on $\5E$,  the wavelets are now parameterized by their \sl location
and scale, \rm just as in standard wavelet analysis \ci{D92}, with the Euclidean time as
scale parameter.

\bull Applying the analytic-signal transform to \eq{RU2} gives
\begin{align}\lab{RU3}
\bt F(z')=\int_\5D  d\m_\5D\0z \ \4W_{z'}^*\4W_z\bt F\0z
=\int_\5D  d\m_\5D\0z\4K(z',\7z)\bt F\0z,
\end{align}
which explains the term ``reproducing kernel.''

\bull By construction, the wavelets $\4W_z$ transform covariantly under the
Poincar\'e  group, which endows their parameters $z$ with physical significance
\ci{K94}. In particular, any resolution \eq{RU} can be boosted, rotated or translated to
give another resolution. Furthermore,  since $\5H$  carries a unitary representation of
the conformal group $\5C$, for which  $\5T$ is a natural domain, it is useful
to study the action of $\5C$ on wavelets. To some extent this has been done in
\ci{K94}.  A connection has been discovered recently with work on  twistor-like
transforms by  Iwo Bilynicki-Birula \ci{B02}, and we have begun a joint project
together with Simonetta Frittelli.

\section{Interpretation as Hertzian pulsed beams}

Recall that the absence of sources was the price we paid at the outset for the
holomorphy used to construct wavelets. But now that we have the wavelets, it  turns
out that we can make them \sl physical \rm (causal!) by introducing sources in a
way that preserves their holomorphy everywhere \sl outside \rm these sources -- 
which is the most that can be expected!  

The introduction of sources will be based on a \sl causal  splitting \rm  of $\2R\0z$
similar to that of  $R\0x$. But whereas the splitting of $R\0x$ is Lorentz-invariant [the
hyperplane $\{t=0\}$ can be tilted without affecting $R\9\pm\0x$], the corresponding
splitting of $\2R\0z$ is frame-dependent.  We must choose a 4-velocity
$v$ ($v^2=1, v_0>0$) and split $z$ into orthogonal temporal and spatial parts
\begin{align*}
z=\t v+z\6S,\qq \t= v z, \qq 
z^2=z_0^2-\3z^2=\t^2+z\6S^2\,.
\end{align*}
Without loss of generality, we take $v=(\30, 1)$ in the following. The general case is
recovered by letting
\begin{align*}
z_0\to\t,\qq  \3z^2\to\t^2-z^2.
\end{align*}
Begin with the factorization
\begin{align*}
z_0^2-\3z^2=(z_0-\2r)(z_0+\2r),
\end{align*}
where 
\begin{align*}
\2r(\3z)=\sr{\3z^2}=\sr{r^2-a^2+2i \3x\cdot\3y},
\qq  \3z=\3x+i\3y,\qq r=|\3x|,\qq a=|\3y|.
\end{align*}
For motivation, think of  $\2r$ as the \sl complex distance \rm from a
``point source'' at $-i\3y$ to the observer at $\3x$. A  study of potential theory based
on complex distances in $\cc n$, and the connections they furnish between elliptic 
and hyperbolic equations, is in progress; see \ci{K00}; see also the related papers
 \ci{K01, K01a, N02}.  Given $\3y\ne\30$, the branch points of $\2r$
form a circle in the plane orthogonal to $\3y$,
\begin{align*}
C(\3y)=\{\3x\in\rr3: r=a,\ \3x\cdot\3y=0\},
\end{align*}
 and following a loop that threads $C$ changes the sign of $\2r$.   We make
$\2r$ single-valued by choosing the ``physical''  branch defined by
\begin{align*}
\re\2r\ge 0, \hb{so that}\3y\to 0\imp \2r\to +r.
\end{align*}

\centerline{ \eps{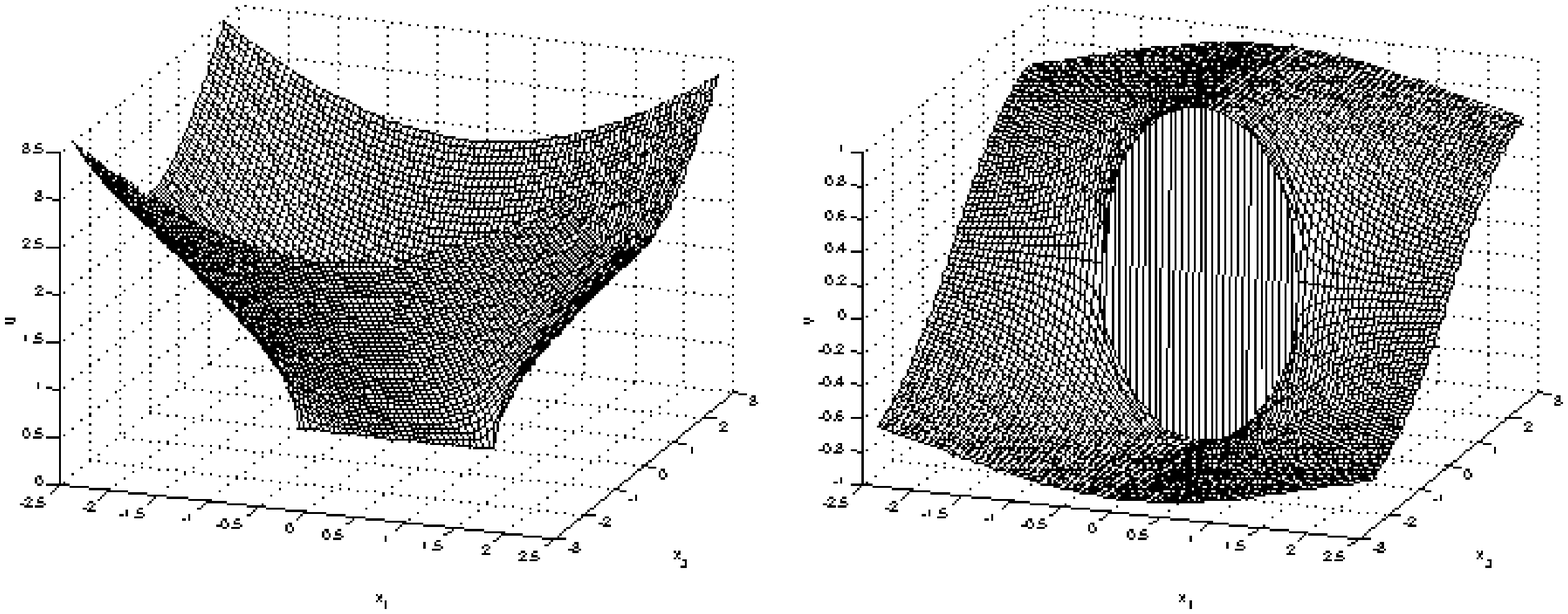}{.55} }

\bf Complex distance: \rm The real part (left) and imaginary part
(right) of $\2r(\3x+i\3y)$  as functions of $\3x=(x_1, 0, x_3)$, with $\3y=(0,0,1)$.
The real part is a pinched cone, with the branch disk in the $x_1$-$x_2$ plane
projected to the interval
$[-1,1]$ of the $x_1$ axis.

\sv2

The resulting branch cut is a disk\footnote{An equivalent branch cut is obtained by
continuously deforming the disk to an arbitrary \sl membrane \rm $M$ 
bounded by $C$.}  bounded by $C$,
\begin{align*}
D(\3y)=\{\3x: r\le a,\ \3x\cdot\3y=0\}.
\end{align*}
 The \sl  extended spatial delta function \rm is defined as the distributional Laplacian
with respect to $\3x$ of the holomorphic Coulomb potential:
\begin{align}\lab{delta}
\2\d^3(\3z)\=-\D \,\frac{1}{4\p\2r(\3z)}.
\end{align}
As a distribution in $\3x$, it is a natural extension of the point source $\d^3(\3x)$.
Roughly speaking, displacing the singularity to $-i\3y$ opens up a ``light cone'' in
$\rr3$ with the $\3y$-axis as ``time,'' of which the branch circle is a ``wave front''
and $D$ its  interior. Similar remarks apply  to extended delta functions in
$\cc n$, which have provided an intriguing connection  \ci{K00} between fundamental
solutions of Laplace's equation in Eulidean $\rr n$ and the initial-value problem for
wave equations in Lorentzian $\rr n$.

Returning to spacetime, $ \2R\0z$ decomposes into partial fractions:
\begin{align}\lab{split}
\2R\0z =\frac{i}{4\p^2(z_0-\2r)(z_0+\2r)}= \2R\9+\0z-\2R\9-\0z,
\end{align}
where
\begin{align}\lab{Rpm}
\2R\9\pm\0z&=\frac{i}{8\p^2\2r(z_0\mp \2r)}\,.
\end{align}
Note that although $\2R\0z=-iS\0z$ is holomorphic in $\5T$, its retarded and
advanced parts $\2R\9\pm\0z$ are \sl not. \rm Viewed on the slice $\4R^4_y$\,, they
are singular on the  world-tube $\2D$ traced out by the branch cut $D$. Everywhere in
$\5T$  outside this \sl source region, \rm they are still holomorphic.

The boundary values of $\2R\9\pm$ are given by the Plemelj formulas
\begin{align*}
\2R\9\pm(x+i0)&=\frac{i}{8\p^2r(t\mp r+i0)}
=\frac{\d(t\mp r)}{8\p r}+\frac{i}{8\p^2 r}\,\5P \frac{1}{t\mp r}\\
\2R\9\pm(x-i0)&
=\frac{i}{8\p^2r(t\mp r-i0)}
=-\frac{\d(t\mp r)}{8\p r}+\frac{i}{8\p^2 r}\,\5P \frac{1}{t\mp r},
\end{align*}
where $\5P$ denotes the Cauchy principal value.  Therefore the jumps across $\rr4$
are
\begin{align}\lab{prop}
\2R\9\pm(x+i0)-\2R\9\pm(x-i0)=\frac{\d(t\mp r)}{4\p r}=R\9\pm\0x.
\end{align}

We now show that $\2R\9\pm\0z$ have very interesting physical interpretations even
when $y\ne 0$,  by looking at their behavior in  slices $\4R^4_y\,$.

Guided by the successful definition \eq{delta} of the extended spatial source,
define the \sl extended spacetime delta function \rm
\begin{align}\lab{delta2}
\2\d^4\0z=\Box \2R\9\pm\0z ,
\end{align}
where $\Box$ is the distributional d'Alembertian acting on $x$.
Since $\Box \2R\9+-\Box \2R\9-=\Box  \2R   =0$, there is no sign ambiguity
on the left. A detailed study of $\2\d^4(x+iy)$ is somewhat involved and will be given
elsewhere \ci{K03}. Here we note the following.

\bull  $\2\d^4(x+iy)$  is a well-defined Schwartz distribution in $\4R^4_y$.

\bull By \eq{prop}, 
\begin{align}\lab{delta3}
\2\d^4(x+i0)-\2\d^4(x-i0)=\Box R\9\pm\0x=\d^4\0x.
\end{align}

\bull  It is easy to show that $\Box\2R\9\pm(x+iy)=0$ at all points of regularity. Thus 
$\2\d^4(x+iy)$ is supported on the world tube $\2D$ representing the evolution of
the source disk in $\4R^4_y$.

\bull In spite of \eq{delta3}, the restrictions of $\2\d^4\0z$ to $\5T\6\pm$  are \sl
not \rm generating functions for $\d^4\0x$ since they are not holomorphic in any
neighborhood of $\rr4$.   (This is what I meant earlier by saying that not all our
constructions fit neatly into hyperfunction theory.)

\bull In the \sl far zone, \rm we have 
\begin{gather*}
r\gg a\imp\2r=\sr{r^2-a^2+2iar\cos\q}\app r+ia\cos\q\\
(z_0=t-is,\ |\3x|=r,\ |\3y|=a,\ \3x\cdot\3y=ra\cos\q),
\end{gather*}
giving a simple expression for the far field from which the pulsed-beam interpretation
can be read off easily:
\begin{align*}
r\gg a\imp  \2R\9\pm\0z\app\frac{1}{8\p^2r}\cdot
\frac{i}{t\mp r+i(s\mp a\cos\q)}\,.
\end{align*}

\centerline{ \eps{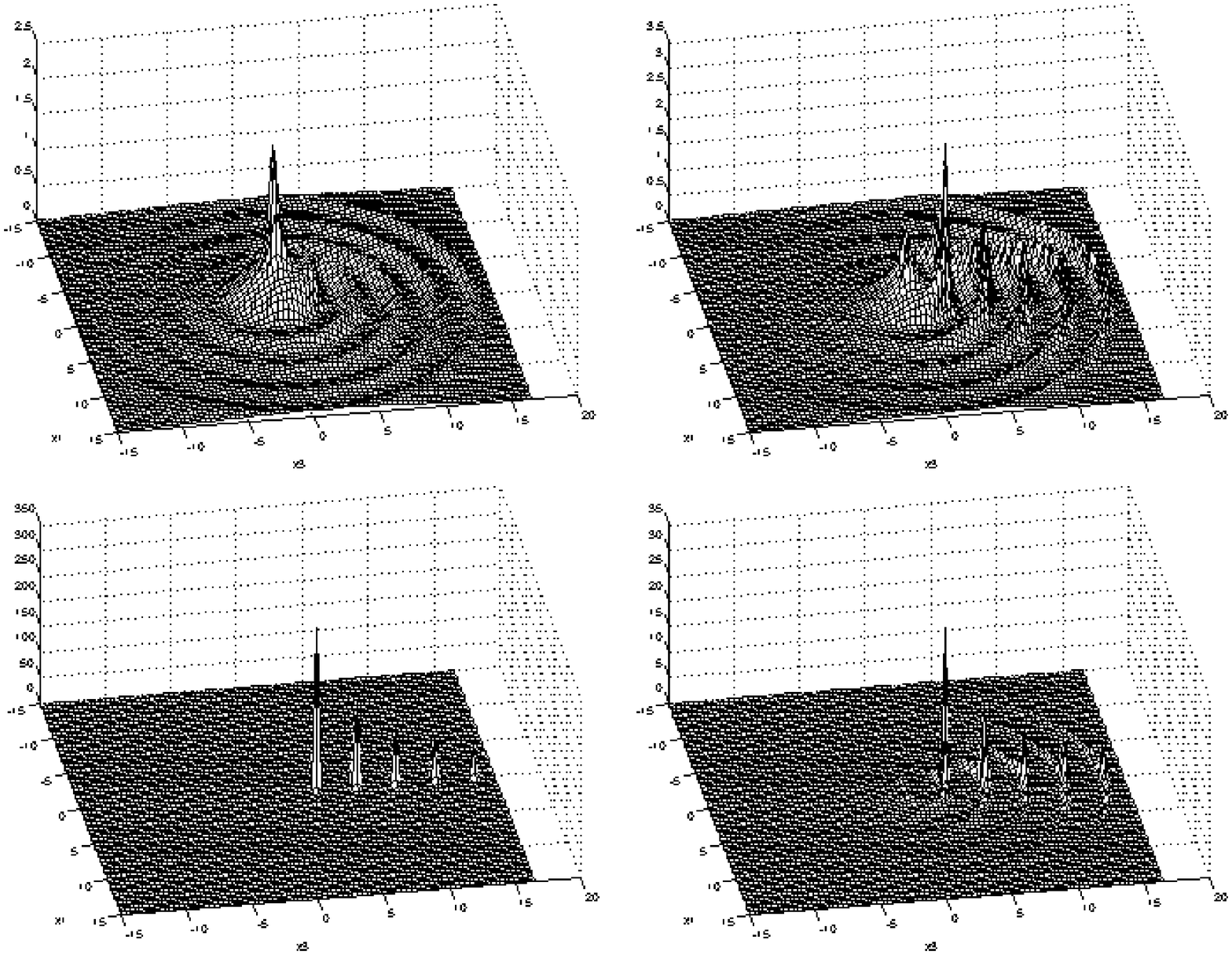}{.6}} 

\bf Time-lapse plots \rm at  $t=1, 5, 10, 15, 20$  of 
$| \2R\9+(x+iy)|$ with $y=(0,0, a, 1)$ as functions of $\3x=(x_1, 0, x_3)$.
\sl Clockwise from upper left: \rm  $a=.5, .9, .99, .999$. The pulses propagate along
$\3y=(0,0,a)$, getting sharper and sharper as $a\to s=1$.

\sv2

\bull $\2R\9-(x+iy)$ is an \sl advanced pulsed  beam \rm converging
toward $D$ along $\3y/s$ and absorbed in  $D$ around $t=0$.

\bull  $\2R\9+(x+iy)$ is a \sl retarded pulsed beam \rm  emitted from $D$ around
$t=0$ and propagating along  $\3y/s$. 

\bull $D$ acts like an \sl antenna dish, \rm simultaneously   absorbing 
$\2R\9-$ and emitting $\2R\9+$, resulting in the sourceless pulsed beam $ \2R(x+iy)$
 focused  at $x=0$.

\bull Both beams have \sl duration \rm $|s|-a$ along the beam axis and are  focused  as
sharply as desired  by letting $(\3y/s)^2\to 1$. As $(\3y/s)^2\to 0$, they become
spherical pulses of duration $|s|$.

Inserting \eq{Rpm} into \eq{Rp} and \eq{Wp} gives 
\begin{align*}
\4W\0z=\4W\9+\0z-\4W\9-\0z,
\end{align*}
where
\begin{align}\lab{Whit}
\4W\9\pm\0z\3p&=i\5L\bt Z\9\pm\0z, \qq  
2\bt Z\9\pm\0z= \2R\9\pm\0z\3p.
\end{align}
By \eq{delta2}, the polarization associated with this Hertz potential is 
\begin{align}\lab{boxZ}
2\bt P\0z=2\Box\bt Z\9\pm\0z=\3p\,\2\d^4\0z.
\end{align}
By \eq{delta3}, the  polarization on $\rr4$ is an \sl impulsive dipole \rm
\begin{align*}
\3P_m\0x-i\3P_e\0x=2\3P\0x=2\bt P(x+i0)-2\bt P(x-i0)=\3p\,\d^4\0x,
\end{align*}
giving an intepretation of $\3p$ in terms of magnetic and electric
dipole moments
\begin{align}\lab{p}
\3p=\3p_m-i\3p_e.
\end{align}
Of course, the extension  $\3P\0x\to\bt P\0z$  mixes the electric and
magnetic dipoles since $\2\d^4\0z$ is complex. Nevertheless, we are free to define 
\sl real \rm  polarizations and causal/anticausal  Hertz potentials and 
fields in the slice
$\4R^4_y$ by
\begin{align*}
\3P_m\0z-i\3P_e\0z&=\3p\,\2\d^4\0z\\
\3Z\9\pm_m\0z-i\3Z\9\pm_e\0z&= \2R\9\pm\0z\3p\\
\3E\9\pm\0z+i\3B\9\pm\0z&=i\5L [\2R\9\pm\0z\3p].
\end{align*}
The left-hand sides are then solutions of Maxwell's equations inheriting the
pulsed-beam interpretations derived for $\2R\9\pm\0z$, but now in the context of  
electrodynamics instead of scalar wave equations.

Whittaker \ci{W04, N55} has shown that given any constant, nonzero vector
$\3v\in\rr3$, a general Hertz potential can be reduced to two scalar potentials   
$\P_e\,, \P_m$ in the form
\begin{align*}
\3Z_e\0x&=\P_e\0x\3v\hb{and}\3Z_m\0x=\P_m\0x\3v, \hb{or}
2 \3Z\0x=(\P_m-i\P_e)\3v.
\end{align*}
Comparison with \eq{Whit} shows that $\2R\9\pm\0z$  are scalar Hertz
potentials of Whittaker type for $2\4W\9\pm\0z\3p$.

Finally, the wavelet decomposition \eq{RU2} of sourceless anti-selfdual fields splits 
into
\begin{align*}
\3F(x')&=\3F\9+(x')-\3F\9-(x'), \qq 
\3F\9\pm(x')=\int_\5D d\m_\5D\0z \4W\9\pm_z(x')\bt F\0z.
\end{align*}
$\3F\9+$ and $\3F\9-$  are fields absorbed and emitted, respectively, by source
disks distributed in $\5D$ according to the coefficient function $\bt F\0z$.
This may be used for analyzing and synthesizing fields with general sources.

Pulsed beams similar to the above have also been studied and applied in the
engineering literature, from a different point of view;  see \ci{HF01} for a recent
review.

\section*{Acknowledgements}

It is a pleasure to thank Iwo Bialynicki-Birula, Ted Newman, Ivor Robinson and Andrzej
Trautman for helpful discussions or suggestions. I also thank Dr.~Arje Nachman for his
sustained support of my work through the Air Force Office of Scientific Research under
Grant  \#F49620-01-1-0271.

\small
\bl {.8}

\end{document}